\documentclass[aps,prd,reprint,showpacs,
superscriptaddress,
nobibnotes,nofootinbib]{revtex4-1}

\usepackage{epsfig}
\usepackage{amssymb} 
\usepackage{amsmath}

\begin{document}

\title{Constraints on Spin-Dependent Short-Range Interaction between Nucleons}

\author{K. Tullney}
\email{Corresponding author: tullnek@uni-mainz.de}
\affiliation{Institut f\"{u}r Physik, Johannes Gutenberg-Universit\"{a}t, 
55099 Mainz, Germany}

\author{F. Allmendinger}
\affiliation{Physikalisches Institut, Universit\"{a}t Heidelberg, 69120 
Heidelberg, Germany}

\author{M. Burghoff}
\affiliation{Physikalisch-Technische Bundesanstalt Berlin, 10587 Berlin,
Germany}

\author{W. Heil}
\affiliation{Institut f\"{u}r Physik, Johannes Gutenberg-Universit\"{a}t, 
55099 Mainz, Germany}

\author{S. Karpuk}
\affiliation{Institut f\"{u}r Physik, Johannes Gutenberg-Universit\"{a}t, 
55099 Mainz, Germany}

\author{W. Kilian} 
\affiliation{Physikalisch-Technische Bundesanstalt Berlin, 10587 Berlin,
Germany}

\author{S. Knappe-Gr\"{u}neberg} 
\affiliation{Physikalisch-Technische Bundesanstalt Berlin, 10587 Berlin,
Germany}

\author{W. M\"{u}ller} 
\affiliation{Physikalisch-Technische Bundesanstalt Berlin, 10587 Berlin,
Germany}

\author{U. Schmidt}
\affiliation{Physikalisches Institut, Universit\"{a}t Heidelberg, 69120 
Heidelberg, Germany}

\author{A. Schnabel} 
\affiliation{Physikalisch-Technische Bundesanstalt Berlin, 10587 Berlin,
Germany}

\author{F. Seifert} 
\affiliation{Physikalisch-Technische Bundesanstalt Berlin, 10587 Berlin,
Germany}

\author{Yu. Sobolev} 
\email{on leave from: PNPI, St.Petersburg, Russia}
\affiliation{Institut f\"{u}r Physik, Johannes Gutenberg-Universit\"{a}t, 
55099 Mainz, Germany}

\author{L. Trahms}
\affiliation{Physikalisch-Technische Bundesanstalt Berlin, 10587 Berlin,
Germany}

\date{\today}

\begin{abstract}
We search for a spin-dependent \textit{P}- and \textit{T}-violating nucleon-nucleon interaction mediated by light pseudoscalar bosons such as axions or axion-like particles. We employed an ultra-sensitive low-field magnetometer based on the detection of free precession of co-located $^{3}$He and $^{129}$Xe nuclear spins using SQUIDs as low-noise magnetic flux detectors. The precession frequency shift in the presence of an unpolarized mass was measured to determine the coupling of pseudoscalar particles to the spin of the bound neutron. For boson masses between 2\:$\mu$eV and 500\:$\mu$eV (force ranges between 3$\cdot10^{-4}$\:m - 10$^{-1}$\:m) we improved the laboratory upper bounds by up to 4 orders of magnitude. 
\end{abstract}


\maketitle

\noindent Axions are light, pseudoscalar particles that arise in theories in which the Peccei-Quinn $U(1)$ symmetry has been introduced to solve the strong \textit{CP} problem \cite{Peccei2}. They could have been created in early stages of the Universe being attractive candidates to the cold dark matter that could compose up to $\sim$1/3 of the ingredients of the Universe \cite{Jaeckel}. Several constraints from astrophysics, cosmology, and  laboratory experiments have been applied in order to prove or rule out the existence of the axion, i.e., constrain the axions mass $m_{a}$ and/or its couplings. The mass range, in which axions are still likely to exist, could thus be narrowed down to a window reaching from $\mu$eV \cite{Asztalos} up to some meV \cite{Raffelt} (axion window).

\noindent Most axion searches look for the conversion of an axion of galactic \cite{Asztalos2}, solar \cite{Arik}, or laboratory \cite{Ehert} origin into a photon in the presence of a static magnetic field. However, any axion or axion-like particle  that couples with both scalar and pseudoscalar vertices to fundamental fermions would also mediate a parity and time-reversal symmetry-violating force between a fermion $f$ and the spin of another fermion $f_{\sigma}$, which is parameterized by a Yukawa-type potential with range $\lambda$ and a monopole-dipole coupling given by \cite{Moody}:
\begin{equation}
	V_{\rm sp}(\vec{r}) = \frac{\hbar^2 g_{s}^{f}g_{p}^{f_{\sigma}}}{8 \pi m_{f_{\sigma}}}(\vec{\sigma} \cdot \hat{r})\left(\frac{1}{\lambda r}+\frac{1}{r^2}\right)e^{-r/\lambda}
\label{eq:potential}
\end{equation}
\noindent $\vec{\sigma}$ is the spin vector and $\lambda$ is the range of the Yukawa-force with $\lambda$= $\hbar/(m_{a}c)$. Thus, the entire axion window can be probed by searching for  spin-dependent short-range forces in the range between 20\:$\mu$m and 0.2\:m. $g_{s}^{f}$ and $g_{p}^{f_{\sigma}}$ are dimensionless scalar and pseudoscalar coupling constants which in our case correspond to the scalar coupling of an axion-like particle to a nucleon ($g_{s}^{f}=g_{s}^{N}$) and its pseudoscalar coupling to a polarized bound neutron ($g_{p}^{f_{\sigma}}=g_{p}^{n}$). Accordingly, we have $m_{f_{\sigma}}=m_{n}$. $\hat{r}$ is the unit distance vector from the bound neutron to the nucleon. The potential given by Eq.\:\ref{eq:potential} effectively acts near the surface of a massive unpolarized sample as a pseudomagnetic field and gives rise to a shift $\Delta\nu_{\rm sp}=2\cdot V_{\Sigma}/h$, e.g., in the precession frequency of nuclear spin-polarized gases ($^{3}$He and $^{129}$Xe), which according to the Schmidt model \cite{Schmidt} can be regarded as an effective probe of spin-polarized bound neutrons. The potential $V_{\Sigma}$ is obtained by integration of $V_{\rm sp}(r)$ from Eq.\:\ref{eq:potential} over the volume of the massive unpolarized sample averaged over the volume of the polarized spin-sample, each having a cylindrical shape. Based on the analytical derivation of $V_{\Sigma,\infty}$ for disc-shaped spin- and matter samples with respective thicknesses $D$ and $d$ \cite{Zimmer}, we obtain
\begin{figure}
\includegraphics[width=3.4in]{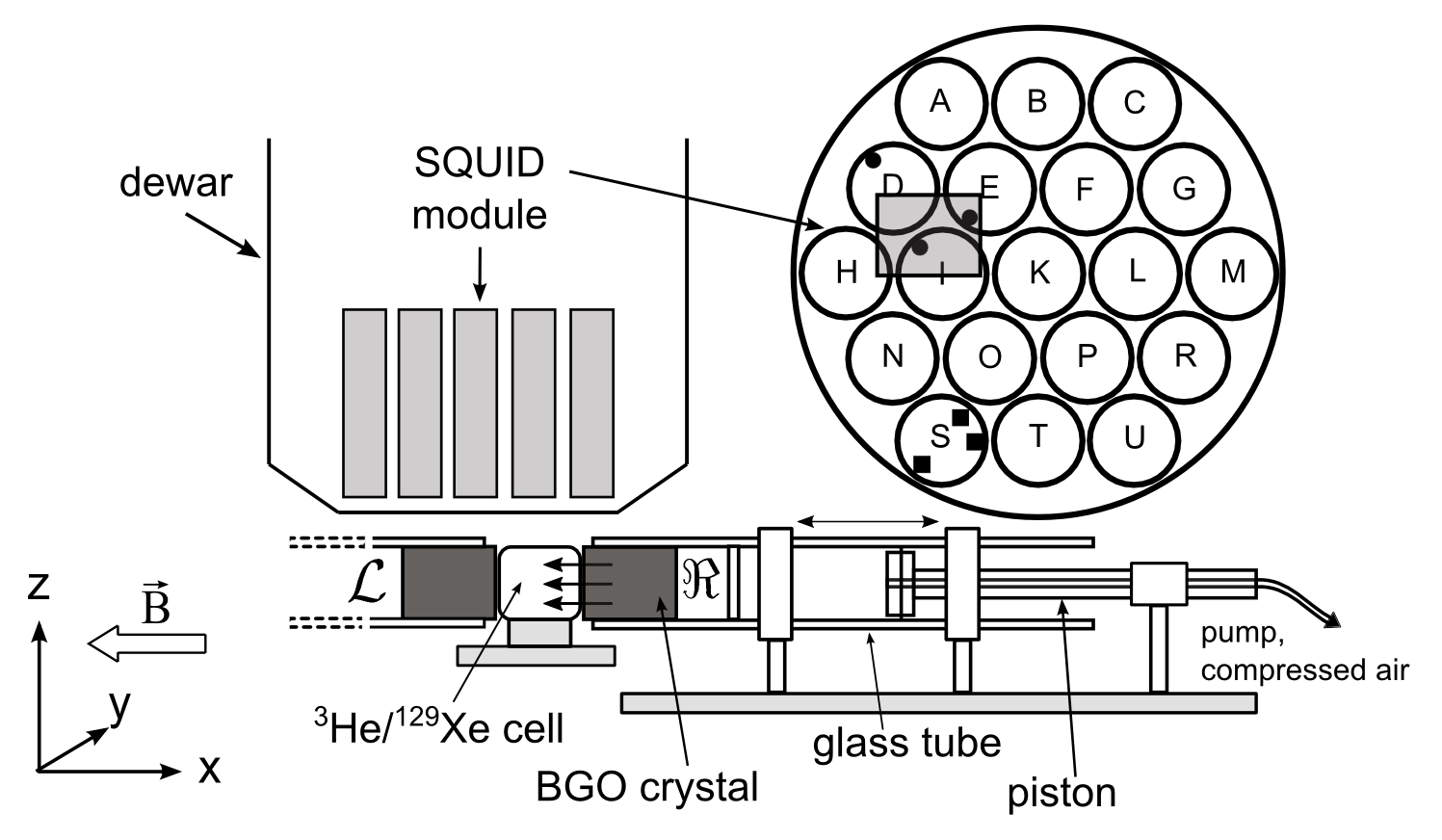}
\caption{Sketch of experimental setup. The lower plane SQUIDs in module D, E, and I marked as ($\bullet$) are used to detect the $^{3}$He/$^{129}$Xe free spin precession. The center of the cylindrical spin sample cell ($D=60$ mm, $\oslash_{D}=58$ mm) has an average distance of $\bar{z}=66$ mm to the sensors. The relative position of the cell in the projection onto the (x,y)-plane is indicated by the grey square. SQUIDs in module S marked as (\tiny{$\blacksquare$}\normalsize) are used for the gradiometric sensor arrangements. The unpolarized mass (cylindrical BGO crystal: $d=70$ mm, $\oslash_{d}=60$ mm) can be moved along the x-axis (\textit{B}-field axis) to "close" ($\Delta x_{c}=2.2\:$mm) and "distant" ($\Delta x_{d}=170\:$mm) position and vice versa (see text). This is accomplished by a piston driven glass tube with the BGO fixed at its cell-facing side. The two measuring arrangements "left" ($\mathcal{L}$) and "right" ($\Re$) are shown.} 
\label{fig:setup}
\end{figure}
\begin{eqnarray}
	V_{\Sigma} &=& V_{\Sigma,\infty} \cdot \eta(\lambda) = 2\pi N \kappa \frac{\lambda^{2}}{D} \cdot e^{-\Delta x/\lambda}\times \nonumber \\
             & & \left(1-e^{-D/\lambda}\right) \cdot \left(1-e^{-d/\lambda}\right) \cdot \eta(\lambda)\:\:.
\label{eq:potentialSigma}
\end{eqnarray}
\noindent $\eta(\lambda)$ takes account for the finite size in transverse direction of our cylindrical samples and $\Delta x$ represents the finite gap between them. Furthermore, $\kappa = \hbar^{2}g_{s}^{N}g_{p}^{n}/(8\pi \cdot m_{n})$ and $N$ is the nucleon number density of the unpolarized matter sample. $\eta(\lambda)$\footnote{$\eta(\lambda)$ can be expressed reasonably well by the fit function $\eta_{\rm fit}(\lambda)=(1+27.8\cdot \lambda^{1.34})/(1+234\cdot \lambda^{1.31})$.} is determined numerically for our cylindrically shaped spin- and matter samples at "\textbf{c}lose"-position (see Fig.\:\ref{fig:setup}).

\noindent Our experimental approach to search for non-magnetic, spin-dependent interactions is to use an ultra-sensitive low-field comagnetometer based on detection of free spin precession of gaseous, nuclear polarized samples \cite{Gemmel}. The Larmor frequencies of $^3$He and $^{129}$Xe in a guiding magnetic field $B$ are given by $\omega_{\rm L,He(Xe)}=\gamma_{\rm He(Xe)}\cdot B$, with $\gamma_{\rm He(Xe)}$ being the gyromagnetic ratios of the respective gas species \cite{Pfeffer, International} with $\gamma_{\rm He}/\gamma_{\rm Xe}=2.75408159(20)$. The influence of the ambient magnetic field and its temporal fluctuations cancels in the difference of measured Larmor frequencies of the co-located spin samples
\begin{equation} 
\Delta\omega=\omega_{\rm He}-\frac{\gamma_{\rm He}}{\gamma_{\rm Xe}}\cdot\omega_{\rm Xe}\:\: .
\label{eq:freqDiff}
\end{equation} 
\noindent On a closer look, a resulting constant frequency shift $\Delta\omega_{\rm lin}$, e.g., due to Earth's rotation, is not compensated by co-magnetometry. That is discussed in \cite{Gemmel2}, together with frequency shifts due to the generalized Ramsey-Bloch-Siegert shift. The latter ones are directly proportional to the particular net magnetization $A_{\rm He(Xe)}\cdot e^{-t/T_{\rm 2,He(Xe)}^*}$ and also are included in the weighted frequency difference $\Delta\omega(t)$: 
\begin{equation}
\Delta\omega(t) = \Delta\omega_{\rm lin} + \epsilon_{\rm He}\cdot A_{\rm He}\cdot e^{-\frac{t}{T_{\rm 2,He}^*}} - \epsilon_{\rm Xe}\cdot A_{\rm Xe}\cdot e^{-\frac{t}{T_{\rm 2,Xe}^*}}\:\: .
\end{equation}	
\noindent Accordingly, its equivalent, the weighted phase difference $\Delta\Phi(t)=\Phi_{\rm He}(t)-\frac{\gamma_{\rm He}}{\gamma_{\rm Xe}}\cdot\Phi_{\rm Xe}(t)$, is sensitive to a phase drift given by 
\begin{eqnarray}
\label{eq:weightedPhase}
\Delta\Phi(t) &=& \Phi_0 + \Delta\omega_{\rm lin}\cdot t - \epsilon_{\rm He}\cdot T_{\rm 2,He}^* \cdot A_{\rm He}\cdot e^{-\frac{t}{T_{\rm 2,He}^*}}\nonumber \\
			  & &  + \epsilon_{\rm Xe}\cdot T_{\rm 2,Xe}^* \cdot A_{\rm Xe}\cdot e^{-\frac{t}{T_{\rm 2,Xe}^*}}\:\: .
\end{eqnarray}
\noindent Due to the knowledge of these side effects, any anomalous frequency shifts generated by non-magnetic spin interactions, such as the quested short range interaction, can be analyzed by looking at $\Delta\omega (t)$ and $\Delta\Phi (t)$, respectively. A sudden frequency change $\Delta\omega_{\rm sp}$ stemming from the pseudoscalar Yukawa potential $V_{\rm sp}(r)$ would occur at an instant $t=t_0$, e.g., by moving a massive matter sample close to the precessing nuclei. This would lead to an additional linear phase drift $\Delta\omega_{\rm sp}\cdot t$ in Eq.\:\ref{eq:weightedPhase} for $t>t_0$. For further analysis, it is useful to develop Eq.\:\ref{eq:weightedPhase} in a Taylor expansion of 5th order\footnote{The criterion to use a Taylor expansion up to the 5th order was that the reduced $\chi^{2}$/d.o.f. of the fit equals 1.} around $t_0$. The weighted phase difference $\Delta\Phi (t)$ can then be described by
\begin{equation} 
\label{eq:poly} 
	\Delta\Phi \left(t'\right)=a+b(t')\cdot t'+c\cdot t'^{2}+d\cdot t'^{3}+e\cdot t'^{4} +f\cdot t'^{5},
\end{equation} 
\noindent with $t' = t-t_0$. The coefficient of the linear term now reads
\begin{equation} 
\label{eq:fitparameters} 
 b(t')  =   \Delta\omega_{\textrm{lin}} +  \Delta\omega_{\rm sp}^{w}(t') + \epsilon_{\rm He}\cdot A'_{\textrm{He}} - \epsilon_{\rm Xe}\cdot A'_{\textrm{Xe}}.
\end{equation}
\noindent Note that $\Delta\omega_{\rm sp}^{w}(t')=2\pi\cdot\Delta\nu_{\rm sp}^{w}\cdot \Theta(\pm t')$ \footnote{($\pm$) in the argument of the Heaviside step function has to be set (-) for the sequence c$\rightarrow$d and (+) for the reverse one d$\rightarrow$c. Furthermore, for runs j = 1, 2, 3 the BGO was moved at $t_{0}$=8700\:s, otherwise at $t_{0}$=10800\:s.} is the only time dependent term in Eq.\:\ref{eq:fitparameters}, so that a change $\delta b=b_{c}-b_{d}=2\pi\cdot\Delta\nu_{\rm sp}^{w}$ of $b(t')$ at $t=t_0$ would directly indicate the existence of the short range interaction. With our special choice of $t' = t - t_0$, the linear coefficient of the Taylor expansion does not dependent on $T_{2}^{*}$ and thus is insensitive to possible changes in $T_{2}^{*}$. The impact of the $T_{2}^{*}$-dependence of higher order terms on the determination of $b(t')$ is discussed in detail in section \textit{systematic uncertainties}.\\

\noindent The experiments were performed inside the magnetically shielded room BMSR-2 at the Physikalisch-Technische Bundesanstalt Berlin (PTB)\cite{Bork}. A homogeneous guiding magnetic field of about 350\:nT was provided inside the shielded room by means of a square coil pair ($B_{x}$-coils) of edge length 1800\:mm. A second square coil pair ($B_{y}$-coils) arranged perpendicular to the $B_{x}$-coils was used to manipulate the sample spins, e.g., $\pi/2$ spin flip by non-adiabatic switching \cite{Gemmel}. The major components of the experimental setup within BMSR-2 are shown in Fig.\:\ref{fig:setup}. For the detection of spin precession we used a multi-channel low-T$_{c}$ DC-SQUID device \cite{Drung, Burghoff}. The $^{3}$He/$^{129}$Xe nuclear spins were polarized outside the shielding by means of optical pumping. Low-relaxation cylindrical glass cells (GE180) were filled with the polarized gases and placed directly beneath the dewar as close as possible to the SQUID sensors. The SQUID sensors detect a sinusoidal change in magnetic flux due to the nuclear spin precession of the gas atoms. In order to obtain a high common mode rejection ratio, three first order gradiometric sensor combinations were used in order to suppress environmental disturbance fields like vibrational modes. Fig.\:\ref{fig:setup} shows their  positions with respect to each other and with respect to the $^{3}$He/$^{129}$Xe sample cell. The system noise of the SQUID gradiometer configurations was between 3 \:fT/$\sqrt{\textrm{Hz}}$ and 10\:fT/$\sqrt{\textrm{Hz}}$ in the range of the $^{3}$He/$^{129}$Xe spin-precession frequencies, i.e., $4\:\textrm{Hz} <\nu_{L}<12\:\textrm{Hz}$, while typical signal amplitudes reached 10\:pT for helium and 3\:pT for xenon at the beginning of the spin precession cycle.
\noindent Typically, the optimum conditions in terms of long transverse relaxation times ($T_{2}^{*}$) and high signal-to-noise ratio (\textit{SNR}) were met at a gas mixture with pressures of $^{3}$He\::\:Xe (91$\%\:^{129}$Xe)\::\:$N_{2}\!\approx\!(2\!:8\!:35)\:$mbar. Nitrogen was added to suppress spin-rotation coupling in bound Xe-Xe van der Waals molecules \cite{Chann, Anger}. As unpolarized matter sample we used a cylindrical BGO crystal (Bi$_{4}$Ge$_{3}$O$_{12}$, $\rho$=7.13 g/cm$^3$). BGO has a high nucleon number density, is a non-conductive material that shows low Johnson-Nyquist noise and is said to have an unusual magnetism-related behaviour in weak constant magnetic fields ($\chi_{\textrm{\textit{mag}}}\approx 0$\:ppm) \cite{Kravchenko, Yamamoto, Grabmaier}. For systematic checks, the BGO crystal could be placed left ($\mathcal{L}$) and right ($\Re$) with respect to the $^{3}$He/$^{129}$Xe sample cell (see Fig.\:\ref{fig:setup}). Since $V_{\textrm{sp}}(\vec{r})\:\propto\: \vec{\sigma}\cdot \hat{r}$, $\Delta\nu_{\textrm{sp}}$ changes its sign in going from $\mathcal{L}$ to $\Re$. This has to be considered by averaging the $\mathcal{L}$ and $\Re$ results. On the other hand, $\Delta\nu_{\textrm{sp}}$ drops out averaging $\mathcal{L}$ and $\Re$ without sign change. In case of a non-zero spin-dependent axion fermion interaction, a shift $\Delta\nu^{w}_{\rm sp}$ in the weighted frequency difference (Eq.\:\ref{eq:freqDiff}) can be extracted from respective frequency measurements in "close" and "distant" position given by
\begin{equation}
\Delta\nu^{w}_{\rm sp} = \frac{2V_{\Sigma}^{c}}{h}\cdot\left(1-\frac{\gamma_{\rm He}}{\gamma_{\rm Xe}}\right),
\label{eq:freqShift}
\end{equation}
\noindent assuming $V_{\Sigma,\rm He}=V_{\Sigma,\rm Xe}=V_{\Sigma,n}\equiv V_{\Sigma}$ (Schmidt model) and $V_{\Sigma}^{d} \ll V_{\Sigma}^{c}$.

\noindent We performed 10 measurement runs lasting approximately 9\:h each. For each measurement run, the BGO crystal was moved after $t_{0}\approx 3\:\textrm{h}$ from "close to distant" position (c$\rightarrow$d) or vice versa (d$\rightarrow$c). The asymmetric timing takes account for the smaller \textit{SNR} in the second measurement block due to the exponential damping ($T_{2}^{*}$) of the signal amplitude which was $T_{\rm 2,He}^{*}\approx 53\:\textrm{h}$ and $T_{\rm 2,Xe}^{*}\approx 5\:\textrm{h}$, typically. By this measure, comparable statistics was obtained for both BGO positions. 

\noindent As discussed in detail in \cite{Gemmel2}, the data from each run were divided into sequential time intervals of $\tau=3.2\:$s. For each obtained sub-data set, a $\chi^{2}$-minimization was performed using an appropriate fit-function to extract the phases $\phi_{\rm He}$, $\phi_{\rm Xe}$ and the frequencies $\omega_{\rm He}$, $\omega_{\rm Xe}$ with the corresponding errors. In a further step, the accumulated phase $\Phi_{\rm He(Xe)}(t')$ was determined for each run in order to derive the weighted phase difference $\Delta \Phi \left(t'\right)$. Then Eq.\:\ref{eq:poly} was fitted simultaneously to the data set $\Delta\Phi (t')$ that was determined for the three gradiometers of each measurement run. From the resulting fit parameters $\overline{a}$, $\overline{b}_{\rm c}$, $\overline{b}_{\rm d}$, $\overline{c}$, $\overline{d}$, $\overline{e}$, $\overline{f}$ and by use of Eqs.\:\ref{eq:fitparameters} and \ref{eq:freqShift}, the frequency shift $\overline{\Delta \nu}_{\rm sp}$ is then extracted from
\begin{equation}	\overline{\Delta\nu}_{\rm sp}=\frac{\bar{b}_{c}-\bar{b}_{d}}{2\pi\cdot (1-\frac{\gamma_{\rm He}}{\gamma_{\rm Xe}})}\:\:.
\label{eq:effectEQ}
\end{equation}
\begin{figure}
\begin{center}
\includegraphics[width=3.4in]{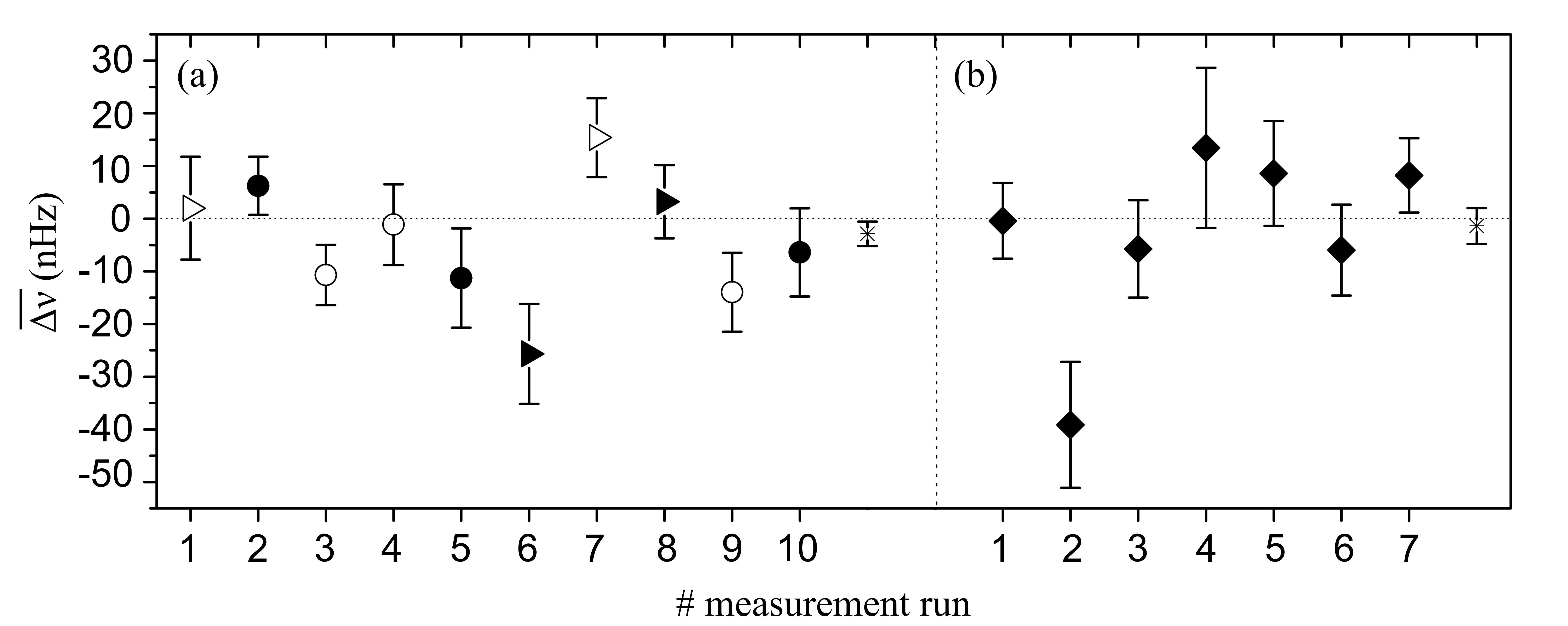}
\caption{(a) Extracted frequency shifts $\overline{\Delta\nu}_{\rm sp}$ (with correlated 1$\sigma$ error) of the 10 measurement runs. The triangles specify the $\Re$, the circles the $\mathcal{L}$ arrangement of the BGO crystal. Full symbols indicate the c$\rightarrow$d sequence, hollow symbols the opposite case (d$\rightarrow$c). (b) Results $\overline{\Delta\nu}_{\rm check}$ obtained from the LV-data using the same fit-model (Eq.\:\ref{eq:poly}). Since no mass was moved, we expect no shift in the spin precession frequency. The rightmost symbols in both plots (stars) indicate the respective weighted means (1$\sigma$ error).} 
\end{center}
\label{fig:PlotData}
\end{figure}		
\noindent For 6 runs ($\#$ : 2, 3, 4, 5, 9, 10), the BGO crystal was positioned at $\mathcal{L}$ otherwise at $\Re$. For all $\mathcal{L}$ runs, the results were multiplied by (-1): $\overline{\Delta\nu}_{\rm sp}=-\overline{\Delta\nu}_{\textrm{sp},\mathcal{L}}$. In Fig.\:2\ref{fig:PlotData}a, values $\overline{\Delta\nu}_{\textrm{sp}}$ for the individual runs are shown together with their correlated 1$\sigma$ errors\footnote{The correlated errors are calculated as square root of the diagonal elements of the covariance matrix of the least $\chi^{2}$ fit-model of Eq.\:\ref{eq:poly} with the proper statistical weights. The uncorrelated errors are about a factor of 30 smaller and not included in the error bars shown in Fig.\:2\ref{fig:PlotData}.}. From the calculation of the weighted mean, one gets $\overline{\overline{\Delta\nu}}_{\rm sp}=(-2.9\pm2.3)\:$nHz. As consistency check, we re-analysed our 2009 data, where we looked for a possible Lorentz-violating (LV) sidereal frequency modulation \cite{Gemmel2}. Since no mass was moved, $\bar{b}_{c}=\bar{b}_{d}$ should hold, using the fit-function of Eq.\:\ref{eq:poly} and a hypothetical time $t_{0}=10800\:\textrm{s}$. Fig.\:2\ref{fig:PlotData}b shows the results $\overline{\Delta\nu}_{\rm check}$ for all 7 measurement runs together with their correlated 1$\sigma$ errors. The weighted mean of the LV-data gives $\overline{\overline{\Delta\nu}}_{\rm check}=(-1.4\pm3.4)\:$nHz. The $\chi^{2}$/d.o.f of the data to their respective weighted means ($\overline{\overline{\Delta\nu}}_{\rm sp}$, $\overline{\overline{\Delta\nu}}_{\rm check}$) gives 2.29 and 2.38, indicating that the errors on the measured frequency shifts (Fig.\:2\ref{fig:PlotData}) are somewhat underestimated. In order to take this into account, the errors are scaled to obtain a $\chi^{2}$/d.o.f of one, as recommended, e.g., by \cite{Particle, Williams}.

\noindent At the 95$\%$ C.L., our results for the measured frequency shifts are:
\begin{eqnarray}
   \label{eq:resultsSP}
   \overline{\overline{\Delta\nu}}_{\rm sp}&=&(-2.9\pm6.9)\:\textrm{nHz}\\
   \overline{\overline{\Delta\nu}}_{\rm check}&=&(-1.4\pm10.5)\:\textrm{nHz}
   \label{eq:resultsCheck}
\end{eqnarray}
\noindent indicating that i) we find no evidence for a pseudoscalar short-range interaction mediated by axion-like particles and ii) the cross check analysis of our LV-data is compatible with zero within the error bars, as expected.\\

\noindent \textit{Discussion of systematic uncertainties:}\\
The movement of the BGO-crystal can produce correlated effects that may mimic a pseudoscalar frequency shift or even compensate the effect we are looking for. Two effects caused by a non-zero magnetic susceptibility of the BGO have to be considered, by taking $\chi_{\rm mag}=-19\:$ppm which is the high field limit ($B >0.1\:$T) \cite{Yamamoto}.

\noindent a.) The BGO at "close" position slightly changes the magnetic field across the volume of the $^{3}$He/$^{129}$Xe sample cell. This effect drops out to first order due to comagnetometry. To second order , however, the difference in their molar masses leads to a difference ($\Delta z$) in their center of masses (barometric formula) which  is $\Delta z = 1.2\cdot 10^{-7}\:\textrm{m}$ for our cylindrical sample cell. This results in a frequency shift of $\Delta\nu_{\rm sys}=\Delta z \cdot \left|\left\langle \partial B/ \partial z\right\rangle\right|_{\rm ind}\cdot \gamma_{\rm He}/2\pi\leq 0.03\:$nHz for induced field gradients in the vertical direction of $\left|\left\langle\partial B/\partial z \right\rangle\right|_{\rm ind}\leq 0.08$\:pT/cm. The field gradients were calculated using \textit{COMSOL Multiphysics}, a finite element analysis software. Compared to the measured frequency shift (Eq.\:\ref{eq:resultsSP}), this systematic effect is negligible.  

\noindent b.) More serious is the fact that a change of the magnetic field gradient by the BGO also influences the $T_{2}^{*}$-times of $^{3}$He and $^{129}$Xe. The direct approach is to extract $T_2^*$ and thus $\Delta T_2^*$ via the exponential decay of the signal amplitudes with the BGO in "close" and "distant" position. The most accurate distinction between $\left(T_{2}^*\right)_{c}$ and $\left(T_{2}^*\right)_{d}$ was obtained through a fit to the amplitude ratio $A_{\rm Xe}(t')/A_{\rm He}(t')$ given by $f_{\rm fit}(t')=W\cdot e^{-t'/T_{\rm eff}^*}$ with $T_{\textrm{eff}}^{*}=T^{*}_{\rm 2,He}\cdot T^{*}_{\rm 2,Xe}/\left(T^{*}_{\rm 2,He}-T^{*}_{\rm 2,Xe}\right)$. According to \cite{Cates, Gemmel, Kilian2}, a relation between $\Delta T^{*}_{\rm 2,He}$, $\Delta T^{*}_{\rm 2,Xe}$, and $\Delta T_{\textrm{eff}}^{*}$ can be derived 
\begin{equation}
  \frac{\Delta T_{\textrm{eff}}^{*}}{\left(T_{\textrm{eff}}^{*}\right)^{2}}=-\frac{\Delta  T_{\rm 2,He}^{*}}{\left(T_{\rm 2,He}^{*}\right)^{2}}+\frac{\Delta  T_{\rm 2,Xe}^{*}}{\left(\rm T_{2,Xe}^{*}\right)^{2}} 
  \approx -0.5\frac{\Delta  T_{\rm 2,He}^{*}}{\left(T_{\rm 2,He}^{*}\right)^{2}}
\label{eq:T2Relation}
\end{equation} 
\noindent by taking the respective diffusion coefficients of $^3$He and $^{129}$Xe in the gas mixture and using the approximation $\oslash_{D}/2=D/2\approx R=30\:$mm for our cylindrically shaped cell. We obtain an upper limit of $\left|\Delta T_{\rm 2,He}^*\right|<160\:$s for a possible $T^{*}_{\rm 2}$--change. From that the systematic frequency shift $\Delta\nu_{\rm sys}^{T_2^*}$ on $b(t')$ due to the higher order terms of (Eq.\:\ref{eq:poly}) can be estimated to be
\begin{equation}
\left|\Delta\nu_{\textrm{sys}}^{T_{2}^{*}}\right| \leq 
\left|\frac{\frac{\Delta T_{\rm 2,He}^*}{\left(T_{\rm 2,He}^*\right)^2}\cdot \left(\frac{E'_{\rm He}}{T_{\rm 2,He}^*}-\frac{1}{2}\frac{E'_{\rm Xe}}{T_{\rm 2,Xe}^*}\right)\cdot \frac{t_0}{2}}{2\pi(1-\gamma_{\textrm{He}}/\gamma_{\textrm{Xe}})}\right|\approx 0.1\:\textrm{nHz}\: .
\label{eq:systemEffect2}
\end{equation}
\noindent Here we used Eq.\:\ref{eq:effectEQ}, replacing $\overline{b}_c$ and $\overline{b}_d$ by the temporal means $2\cdot \overline{c}_c \cdot\left\langle t'\right\rangle_{t'}$ and $2\cdot \overline{c}_d \cdot\left\langle t'\right\rangle_{t'}$ of the quadratic term in Eq.\:\ref{eq:poly} with $\overline{c}_{c(d)}=\left\{-(E'_{\rm He}/2)/(T_{\rm 2,He}^*)^2+(E'_{\rm Xe}/2)/(T_{\rm 2,Xe}^*)^2\right\}_{c(d)}$ and $E'_{\rm He(Xe)}=\epsilon_{\rm He(Xe)}\cdot A'_{\rm He(Xe)}\cdot T_{\rm 2,He(Xe)}^*$. Values for the respective $E'_{\textrm{He(Xe)}}$ phase amplitudes were extracted from the fit function (Eq.\:\ref{eq:poly}) applied to the data and result to be $\left\langle E'_{\textrm{He}}\right\rangle =11.5\:$ rad and $\left\langle E'_{\textrm{Xe}}\right\rangle = 0.1\:$rad. Finally, $\left\langle t' \right\rangle_{t'}$ was taken to be $\left\langle t' \right\rangle_{t'}\approx t_0/2$. 

\noindent From Eq.\:\ref{eq:systemEffect2} a conservative estimate of the systematic error can be made with $\vert\Delta\nu_{\rm sys}^{T_{2}^{*}}\vert = \pm 0.2\:$nHz (95$\%$ C.L.), which brings us to the final result
\begin{equation}
\overline{\overline{\Delta\nu}}_{\textrm{sp}}=(-2.9 \pm 6.9 \pm 0.2)\:\textrm{nHz} \:\:\:(\textrm{95\% C.L.})
\end{equation}

\noindent for the measured pseudoscalar frequency shift. 

\begin{figure}		
\includegraphics[width=3.45in]{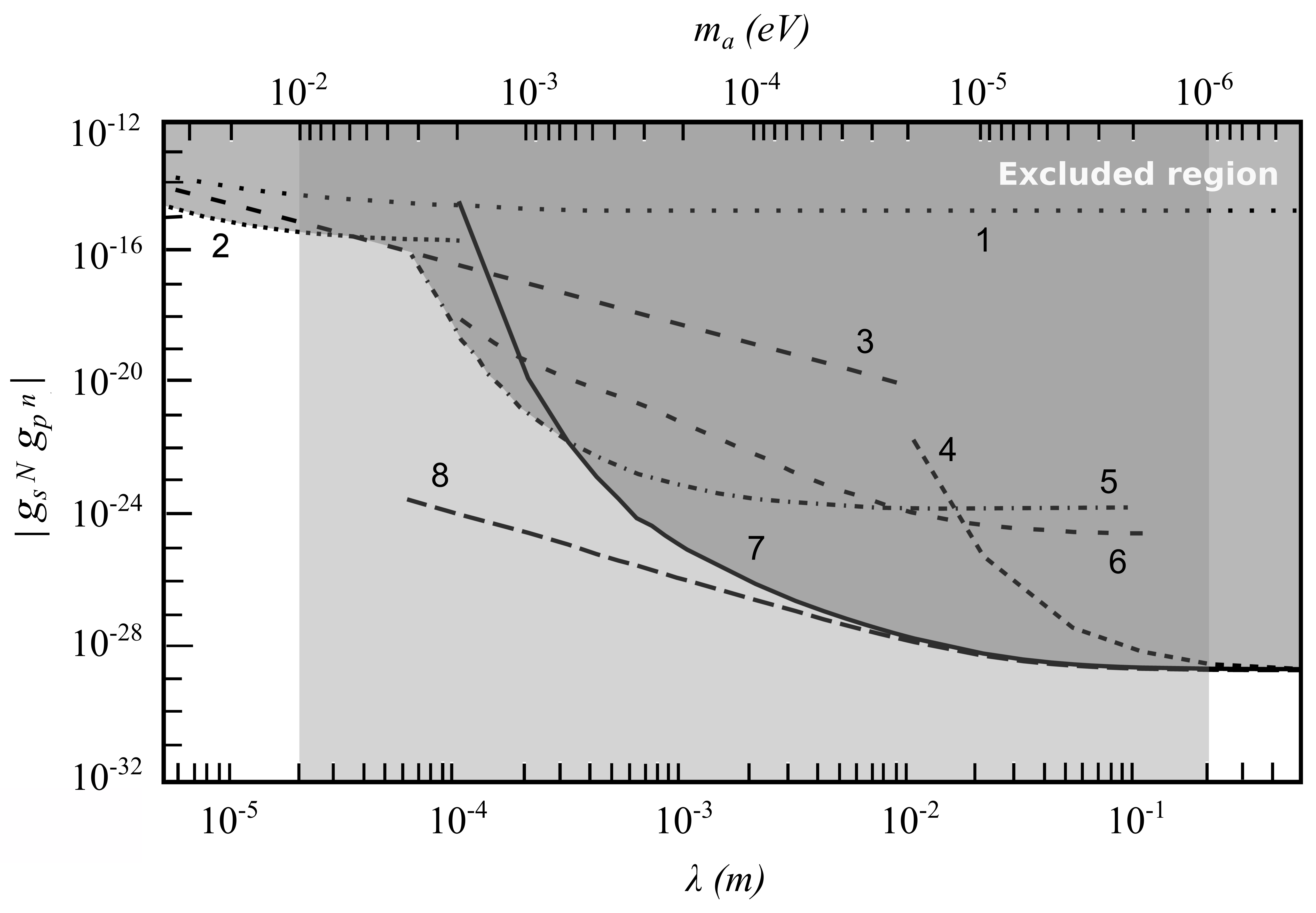} 
\caption{The experimental 95$\%$ confidence upper limit on $\left|g_{s}^{N}g_{p}^{n}\right|$ plotted versus $\lambda$, the range of the Yukawa-force with $\lambda=\hbar/(m_{a}c)$. The axion window is indicated by the light grey area. (1): result of \cite{Baessler}, (2): result of \cite{JenkeNeu}, (3): result of \cite{SerebrovNeu}, (4): result of \cite{Petukhov}, (5): result of \cite{Youdin}, (6): result of \cite{Bulato}, (7): result of \cite{Chu}, (8): this experiment ($\Delta x=2.2\:$mm) \cite{note}, (9): expected results for $\Delta$x$\approx$ 0\:mm using the same data set demonstrates the gain in measurement sensitivity for $\lambda<10^{-3}$\:m. See \cite{Hoedel} for bounds on the pseudoscalar short-range force between polarized electrons and unpolarized nucleons. Raffelt \cite{Raffelt2} points out that much tighter constraints on $\left|g_{s}^{N}g_{p}^{n}\right|$ can be inferred by combining constraints on $g_{s}$ from stellar energy-loss arguments and $g_{p}$ from searches for anomalous monopole-monopole forces.} 
\label{fig:plot}
\end{figure}

\noindent From the total error $\delta(\overline{\overline{\Delta\nu}}_{\textrm{sp}})=\pm 7.1$\:nHz we can then derive exclusion bounds for $\left|g_{s}^{N}g_{p}^{n}\right|$ using Eq.\:\ref{eq:potentialSigma} and $\vert\delta(\overline{\overline{\Delta\nu}}_{\textrm{sp}})\vert \geq 2\cdot V_{\Sigma}^{c}/h$ which are shown in Fig.\ref{fig:plot}. 

\noindent We have substantially improved the bounds on a spin-dependent short-range interaction  between polarized (bound) neutrons and unpolarized nucleons over most of the axion window, tightening existing constrains on axion-like particles heavier than 20 $\mu$eV by up to four orders of magnitudes. 

\noindent And there are clear strategies on how to improve our experimental sensitivity:
i) Close contact of the spin system with the matter sample. For $\Delta x\approx 0$\:mm, our present measurement senitivity will significantly increase for $\lambda < 10^{-3}$\:m (see Fig.\ref{fig:plot}).
ii) Moving the spin/matter sample more frequently between its set positions (c$\leftrightarrow$d and/or $\mathcal{L}\leftrightarrow\Re$). This results in a different time structure for the linear term in the fit model of Eq. \ref{eq:poly} such that the correlated error approaches the uncorrelated one. This was demonstrated in \cite{Gemmel2}, already. 
iii.) Magnetic susceptibility related artefacts have to be eliminated by taking zero-susceptibility matched matter samples ($\chi_{\textrm{mag}}\approx$ 0\:ppm) as it is common practice in high resolution NMR spectroscopy \cite{Ravi}. 

\noindent This work was supported by the Deutsche Forschungsgemeinschaft (DFG) under contract number BA 3605/1-1 and the research center "Elementary Forces and Mathematical Foundations" (EMG) of the University in Mainz, and by PRISMA cluster of excellence at Mainz. We are grateful to our glass blower R. Jera for preparing the low relaxation glass cells from GE180.

\end{document}